# Image encryption schemes for JPEG and GIF formats based on 3D baker with compound chaotic sequence generator

Shiyu Ji, Xiaojun Tong, Miao Zhang

School of computer science and Technology, Harbin Institute of Technology, Weihai, 264209;China

E-mail: jishiyu1989@google.com, tong_xiaojun@163.com, zhangmiaozm209@126.com

## Abstract

This paper proposed several methods to transplant the compound chaotic image encryption scheme with permutation based on 3D baker into image formats as Joint Photographic Experts Group (JPEG) and Graphics Interchange Format (GIF). The new method averts the lossy Discrete Cosine Transform and quantization and can encrypt and decrypt JPEG images lossless. Our proposed method for GIF keeps the property of animation successfully. The security test results indicate the proposed methods have high security. Since JPEG and GIF image formats are popular contemporarily, this paper shows that the prospect of chaotic image encryption is promising.

## 1. Introduction

For the past several decades, we have witnessed the rapid development of chaos theory and practices, as well as the astounding growth in the demand of transmitting images via the Internet. To reduce the flow rate of image file transmission, a lot of research on image compression has been carried out [1-3]. Joint Photographic Experts Group (JPEG) is a successful image format standard. In JPEG, mathematical tools like discrete cosine transformation (DCT) and quantization are introduced when processing the original image data [4]. JPEG has a great performance on reducing the image file size. However, since DCT and quantization are both lossy transformations, it is difficult to encrypt and decrypt JPEG image files lossless applying the general cryptographic scheme within the scope of bitmap images. There are several alternatives on JPEG image encryption [5-6], but none of them will produce cipher images, the pixel data of which owns a satisfied randomness.

Fortunately, it is not unpractical to apply the chaotic image encryption to JPEG, as both of the lossy transformations mentioned above can be averted in our proposed scheme. Therefore, we obtain a lossless cryptographic scheme that is practical to JPEG images.

Graphics Interchange Format (GIF) is another distinguished image format, since it has high compression ratio and animate frames [7]. Though GIF has the same RGB color space as normal bitmaps, the palette is a new case and it needs encryption as well. There are limited papers about encryption for these features now, and this paper will offer some new solutions.

Compound chaotic sequence generator based on 3D baker is an excellent stream cipher scheme for bitmaps, since it has a big enough key space, and a long sequence period due to the introduction of perturbations onto the neighborhood of the fixed points, and high security because of the combination of diffusion and permutation [8-9]. Please note that our proposed scheme is an extension of it, expanding the scope of images to be dealt with.

The rest of this paper is organized as follows. In Section 2, there is a brief introduction to JPEG and GIF image file formats, and a discussion about compound chaotic sequence generator and permutation based on 3D baker map. The proposed methods applying the chaotic image encryption schemes to JPEG and GIF reside in Section 3 and finally the security test results and analysis of our new schemes are in Section 4, with NIST SP-800 22 test included [10].

## 2. Research on JPEG and GIF file formats

*2.1 Joint Photographic Experts Group (JPEG) file format*

According to the international specification, a JPEG file can be encoded in various ways. However, most implementations adopt JFIF encoding [4]. Therefore, we have a discussion about JFIF encoding as follows.

The image data is stored into three components:

(1). Y: luminance component, standing for brightness;

(2). Cb: blue-difference chromatic component;

(3). Cr: red-difference chromatic component.

Each component is split into several 8×8 pixels, where the data undergoes a discrete cosine transform and quantization [5].

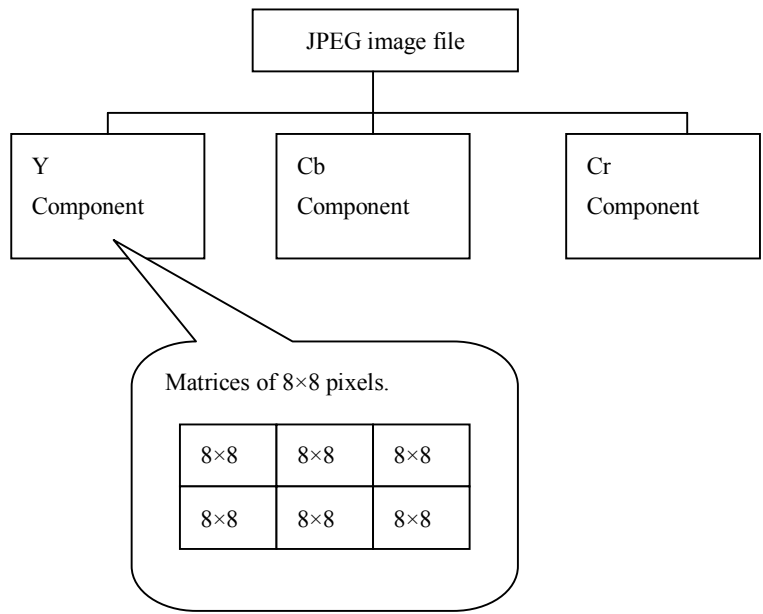

Fig. 1 JPEG image file structure

In order to get access to all the matrices in one component, we need introduce two significant terms: sampling factor and minimum coded unit (MCU).

(1) Sampling factor

Because of the distribution of color-sensitive and brightness-sensitive receptors in the human eyes, people can see much more accurately in the brightness of a scene (the luminance component) than in the hue and color saturation (the chromatic components) of it. Therefore, we can design image encoders that compress images more efficiently based on this fact.

The transformation into the YCbCr color model enables the next step, which is to reduce the spatial resolution of the Cb and Cr components. This step is called downsampling or chroma sub sampling. The most used ratios for downsampling in JPEG images are 4:4:4 (no downsampling), 4:2:2 (reduction by a factor of 2 in the horizontal direction), or most commonly, 4:2:0 (reduction by a factor of 2 in both the horizontal and vertical directions).

Here we give the definition of sampling factor:

*Definition 1*: For each component, sampling factors $H_i$ and $V_i$ are defined relating component dimensions $x_i$ and $y_i$ to maximum dimensions $X$ and $Y$, according to the following expressions:

$$x_i = \left\lceil X \times \frac{H_i}{H_{max}} \right\rceil, y_i = \left\lceil Y \times \frac{V_i}{V_{max}} \right\rceil \qquad (1)$$

where $H_{max}$ and $V_{max}$ are the maximum sampling factors for all components in the frame, and $\lceil \ \rceil$ is the ceiling function [4].

Based on the specification, the sampling factors also specify the number of horizontal or vertical data units in each corresponding component. The JPEG standard declares it in Frame Header syntax.

(2) Minimum coded unit (MCU)

MCU is the sequence of the data units defining by the sample factors of the component in the scan. For interleaved order, which is the most common situation among various implementations, each component is partitioned into small rectangular arrays of $H_k$ horizontal data units by $V_k$ vertical data units. Here $H_k$ and $V_k$ denote the horizontal and vertical sampling factors.

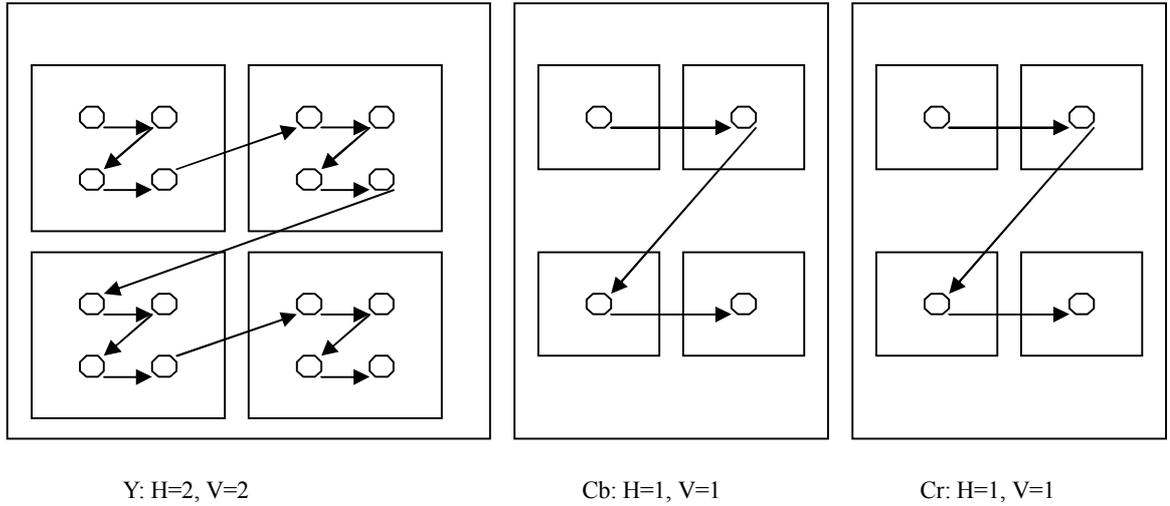

Y: H=2, V=2            Cb: H=1, V=1            Cr: H=1, V=1

Fig. 2 Interleaved data order example

Fig. 2 indicates the most common situation (4:2:0) of interleaved data. Then the MCU can be written as follows.

$$\begin{aligned} MCU_1 &= d_{00}^Y d_{01}^Y d_{10}^Y d_{11}^Y d_{00}^{Cb} d_{00}^{Cr} \\ MCU_2 &= d_{02}^Y d_{03}^Y d_{12}^Y d_{13}^Y d_{01}^{Cb} d_{01}^{Cr} \\ MCU_3 &= d_{20}^Y d_{21}^Y d_{30}^Y d_{31}^Y d_{10}^{Cb} d_{10}^{Cr} \\ MCU_4 &= d_{22}^Y d_{23}^Y d_{32}^Y d_{33}^Y d_{11}^{Cb} d_{11}^{Cr} \end{aligned} \qquad (2)$$

In formula (2), $d_{ij}^Y$ stands for the matrix of 8×8 pixels in i-th row and j-th column of Y component.

Therefore, we have the formula to work out the height and width of the particular component in a JPEG image.

$$\begin{aligned} Width &= H \times X \times 8 \\ Height &= V \times Y \times 8 \end{aligned} \qquad (3)$$

Here $H$ and $V$ are horizontal and vertical sampling factors. $Y$ stands for number of lines, and $X$ stands for number of MCUs per line. Both of them are specified in the frame header.

The three tables work as the basis of our encryption scheme.

*2.2 Graphics Interchange Format (GIF) file format*

A GIF image file has a global palette and several frames and local palettes, if exist. In each frame, the image data is stored in the form of indices to the global (or local) palette and the data is compressed by Lempel-Ziv-Welch (LZW) algorithm [7][11].

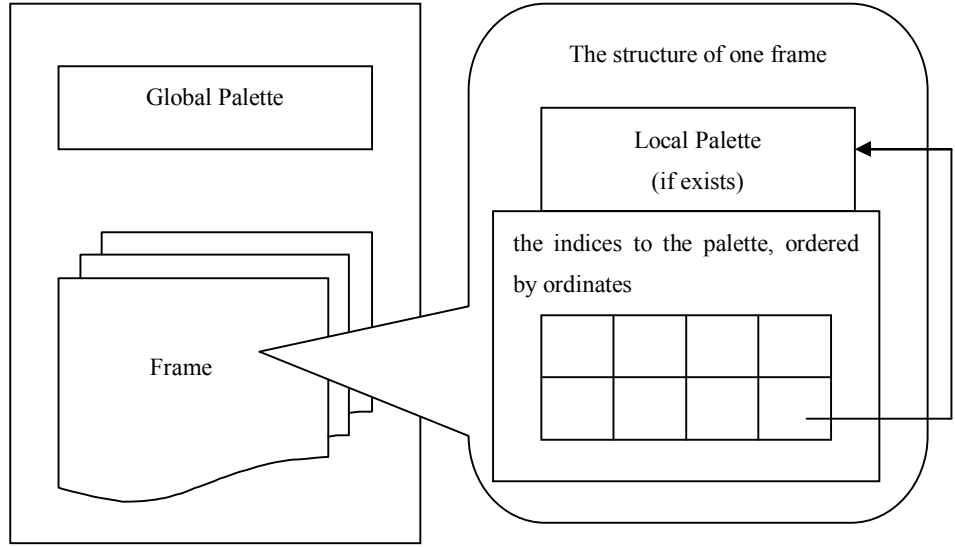

Fig. 3 the architecture of GIF file

The indices of each frame is assigned by horizontal and vertical ordinates, just as the same as normal bitmaps. However, to reduce the file size totally, each frame only describes part of the logical screen, which means the indices do not start from the origin (0,0) every time. The left and top ordinates of each frame, as well as width and height, are specified in the Image Descriptor.

*2.3 Compound chaotic sequence generator*

Chaotic sequence is generated by a sequence generator, which calculates a chaotic function by iterations. The compound function used to generate the sequence is shown below.

$$\begin{cases} f_0(x_{n-1}) = 4x_{n-1}^3 - 3x_{n-1} \\ f_1(y_{n-1}) = 8y_{n-1}^4 - 8y_{n-1}^2 + 1 \\ z_{k_0+k_1+1} = \begin{cases} x_{k_0+1} = f_0(x_{k_0}), x_{k_0} + y_{k_1} < 0 \\ y_{k_1+1} = f_1(y_{k_1}), x_{k_0} + y_{k_1} \geq 0 \end{cases} \end{cases} \quad (4)$$

where $x, y \in I = [-1, 1]$

It is proved that formula (4) is chaotic and we can use it to generate chaotic sequence as follows.

$$S_1(k) = \begin{cases} \left\lfloor (1 - \frac{\arccos(z_k)}{\pi}) \cdot N \right\rfloor, z_k \in [-1, 1) \\ N - 1, z_k = 1 \end{cases} \quad (5)$$

where $N$ stands for the number of sub domains, which form bit sequence with certain length.

Small perturbations should be applied within the neighborhood of the fixed point, as computer has limited precision of floating computations.

*2.4 Permutation based on 3D baker*

In order to apply 3D baker, the image data should be converted into a unit cube.

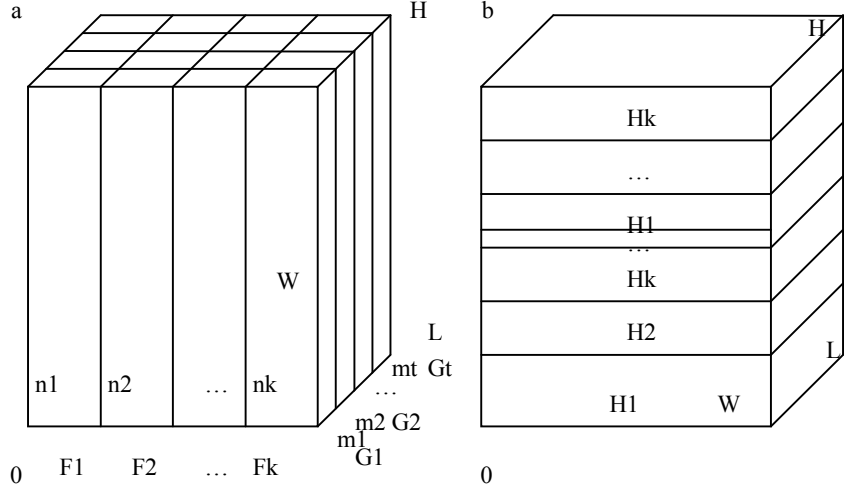

Fig. 4 (a) 3D baker map dividing and (b) 3D baker

As what Fig. 4 shows, the unit cube is furthermore divided into many tinier blocks. The total number of the blocks is $k \times t$ in Fig. 4.

To define the width, length and height of the cube, we apply the formula (6) as follows.

$$\begin{cases} W = \dfrac{M}{4}, L = \dfrac{N}{8}, H = 32, N \bmod 8 = 0 \\ W = \dfrac{M}{4}, L = \dfrac{N}{4}, H = 16, N \bmod 8 \neq 0, N \bmod 4 = 0 \\ W = \dfrac{M}{4}, L = \dfrac{N}{2}, H = 8, N \bmod 4 \neq 0, N \bmod 2 = 0 \\ W = \dfrac{M}{4}, L = N, H = 4, N \bmod 2 \neq 0 \end{cases} \quad (6)$$

where $M$ denotes the vertical dimensions of the array to be processed, and $N$ represents the horizontal dimensions of the array.

After that, the number of blocks should be determined. The number $k$ and $t$ should be initialized and they have to satisfy the conditions:

$$k < W, t < L \quad (7)$$

As Fig. 5, we apply the compound chaotic sequence to determine the width and length of each block in Fig. 4. We obtain the value the width or length using formula (8).

$$\begin{aligned} W_i &= \left\lfloor \dfrac{(w_i + 1)}{2} \cdot W \right\rfloor \\ L_i &= \left\lfloor \dfrac{(l_i + 1)}{2} \cdot L \right\rfloor \end{aligned} \quad (8)$$

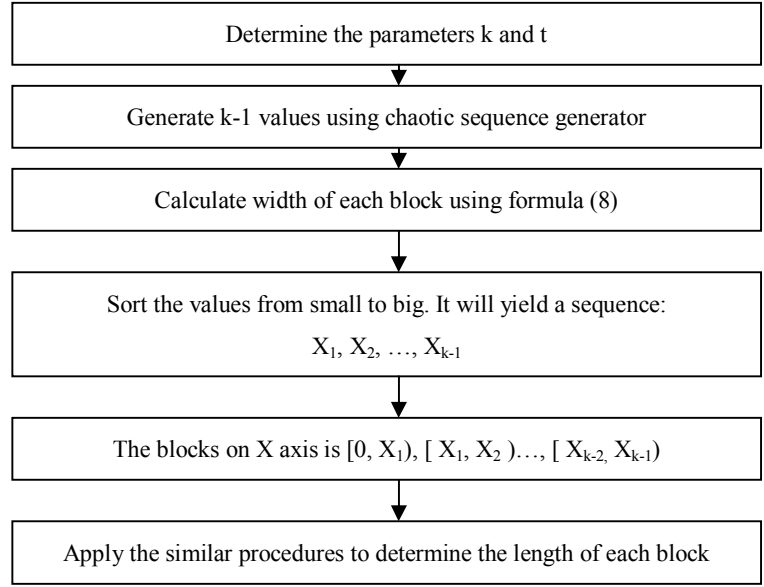

Fig. 5 The procedures of x-axis and y-axis dividing

The 3D baker map is as formula (9).

$$B_3(x,y,z) = (\mod(\mod(num,WL),W), \left\lfloor \frac{\mod(num,WL)}{W} \right\rfloor, \left\lfloor \frac{num}{WL} \right\rfloor)$$
$$num = (WG_j + m_j F_i)H + zm_j n_i + (y - G_j)n_i + x - F_i \quad (9)$$
$$F_i = \sum_{k=1}^{i-1} n_k, F_1 = 0, G_j = \sum_{k=1}^{j-1} m_k, G_1 = 0$$

# 3. Compound chaotic image encryption based on 3D baker for JPEG and GIF files

*3.1 JPEG chaotic encryption scheme*

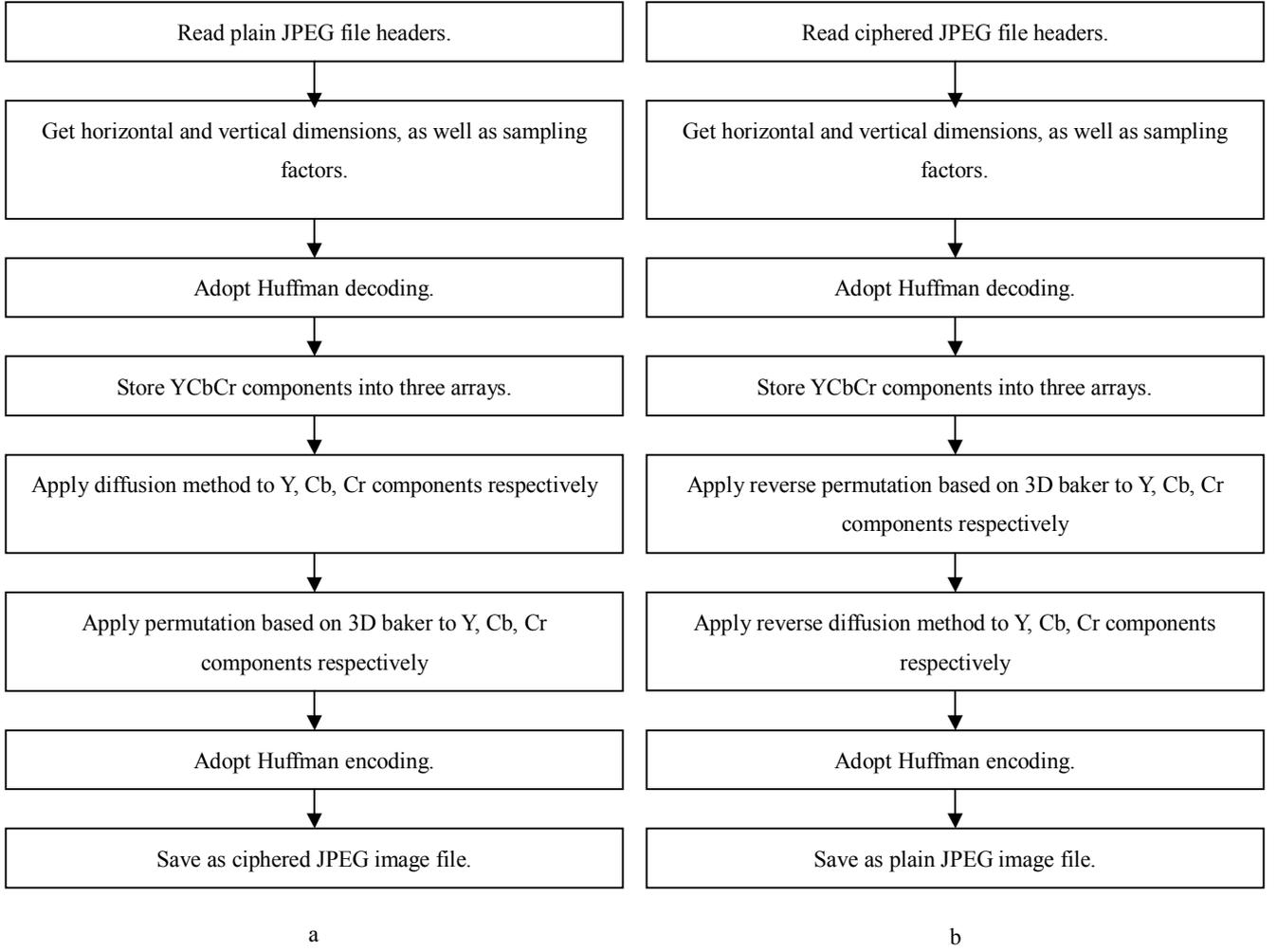

Fig. 6 (a) procedures of JPEG encryption and (b) procedures of JPEG decryption

Fig. 6 shows the main procedures of encryption and decryption in our proposed scheme. As you can see, our method avoids the discrete cosine transform (DCT) and quantization and implement a lossless encryption and decryption scheme on JPEG images. The methods of diffusion and the permutation need further explanations, which are given as follows.

*(1) Diffusion*

According to section 2.1, in MCU, each matrix consists of 64 values, which are signed words. Most of the values range from -128 to 127. Therefore, we have to leave the most significant bit of each value and choose a Huffman table which support long DC values (more than 8 bits) to make sure the subsequent encoding will not receive any exceptions. For diffusion of encryption, we apply the normal compound chaotic sequence, which has been introduced in section 2.3, to this new situation as Formula (10).

$$C_0 = I_0^{(7\ldots15)} \oplus (((S_1(0) + I_0^{(0\ldots6)}) \bmod 128) \oplus S_1(n))$$
$$C_i = I_i^{(7\ldots15)} \oplus (((S_1(i) + I_i^{(0\ldots6)}) \bmod 128) \oplus C_{i-1}^{(0\ldots6)})$$

(10)

We give a brief explanation to the symbols mentioned in Formula (10). $C_i$ means the i-th word of ciphered text. $S_1(i)$ stands for the i-th byte of the chaotic sequence generated by Formula (4) and

(5). $I_i$ represents the i-th word of plaintext. $n$ represents the total length of the plaintext, and the $S_1(n)$ is the particular byte generated for the first element. The $I_i^{(j...k)}$ means the i-th signed word in plaintext, with its j-th to k-th bits preserved, and other bits set to 0. The order of the words is from left to right, row by row, in a single matrix, and the order is the same as what Fig. 2 illustrates among different matrices.

The diffusion of the decryption is as Formula (11).

$$\begin{aligned} I_i &= C_i^{(7...15)} + (C_{i-1}^{(0...6)} \oplus C_i^{(0...6)} + 128 - S_1(i)) \bmod 128 \\ I_0 &= C_0^{(7...15)} + (S_1(n) \oplus C_0^{(0...6)} + 128 - S_1(0)) \bmod 128 \end{aligned} \quad (11)$$

It is recommended to take this diffusion method bilaterally, which means the image is encrypted both forwards and backwards. Such diffusion will obstruct the chosen plaintext attack.

*(2) Permutation*

According to Formula (6), the width, length and height are determined in different conditions of the measurement of the image. In Formula (6), the $M$ and $N$ are obtained as follows.

$$M = V \times Y, N = H \times X \quad (12)$$

Here $H$ and $V$ are horizontal and vertical sampling factors. $Y$ stands for number of lines, and $X$ stands for number of MCUs per line. The smallest unit of permutation is 8x8 matrix of pixels.

We apply the 3D baker map to component Y, Cb and Cr respectively.

*3.2 GIF chaotic encryption scheme*

Fig. 7 represents the main procedures of GIF image encryption in our proposed scheme. The permutation procedure is the same as that of the normal bitmaps. Therefore, we only have to illustrate the diffusion procedure here.

The diffusion of encryption for image data is the same as the scheme for normal bitmaps (Formula (13)).

$$\begin{aligned} C_0 &= ((S_1(0) + I_0) \bmod N) \oplus S_1(n) \\ C_i &= ((S_1(i) + I_i) \bmod N) \oplus C_{i-1} \end{aligned} \quad (13)$$

The definitions of the symbols that occur in Formula (13) remain the same as those in Formula (10).

The diffusion of decryption is shown as follows.

$$\begin{aligned} I_i &= (C_{i-1} \oplus C_i + N - S_1(i)) \bmod N \\ I_0 &= (S_1(n) \oplus C_0 + N - S_1(0)) \bmod N \end{aligned} \quad (14)$$

We apply the same diffusion solution to both global palette and local palette. It will bring more randomness to the ciphered image.

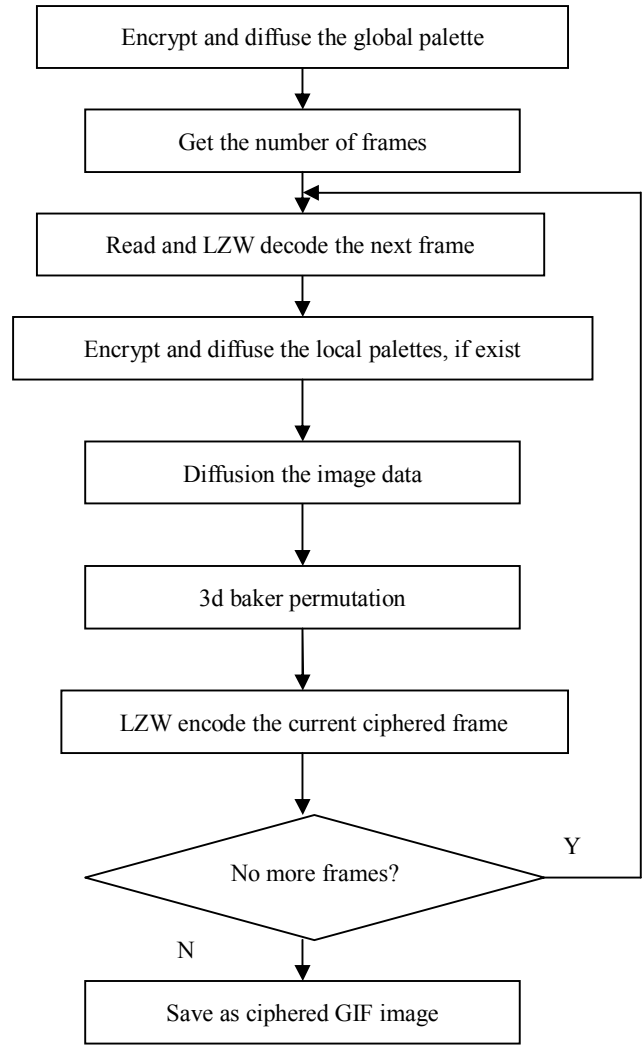

Fig. 7 Encryption of GIF images

The experiment results show that our scheme can keep the property of animation, which is a novel feature of GIF images. The frames are dealt with individually, and therefore, they will not affect each other. The animation of the ciphered image shows a good property of randomness.

## 4. Security analysis

For the two new cryptographic schemes, we conducted two groups of security tests.
The main parameters and conditions are given as follows:
(1) For JPEG encryption, we adopt flower. jpg as original image ( Fig. 8(a) ). The horizontal and vertical sums k and t are 32 and 20 respectively during the 3D baker map. The diffusion is bilateral, both forwards and backwards.
(2) For GIF encryption, we adopt chain. gif as original image ( Fig. 9(a) ). The horizontal and vertical sums k and t are 5 and 6 respectively during the 3D baker map. The diffusion is bilateral, both forwards and backwards.

There are two traditional image encryption solutions available for comparison: 3DES and

AES128. The plaint image data is encrypted by 3DES and AES128 in blocks under electronic codebook scheme.

## 4.1 Space of the key

If the precision of both initial values is $10^{-14}$, the size of the key space is $4 \times 10^{28}$. Combined with the size of the space of confusion, which is $W \times L \times n$, where $n$ stands for the times of iteration, the total size of the key space of our cryptographic system is $W \times L \times n \times 4 \times 10^{28}$, eliminating five fixed points and some singularities. This key space is big enough to resist brute force attack.

## 4.2 Correlation of two adjacent pixels

The calculation of correlation is as follows.

$$C_r = \frac{N\sum_{j=1}^{N}(x_j y_j) - (\sum_{j=1}^{N} x_j)(\sum_{j=1}^{N} y_j)}{\sqrt{(N\sum_{j=1}^{N} x_j^2 - (\sum_{j=1}^{N} x_j)^2)(N\sum_{j=1}^{N} y_j^2 - (\sum_{j=1}^{N} y_j)^2)}} \quad (15)$$

The test results are shown in Table 1. The initial values are -0.1790288311 and -0.1628589871.

Table 1 Correlation coefficient of adjacent pixels in two images

| JPEG Images | | Orientation | | |
|---|---|---|---|---|
| | | Horizontal | Vertical | Diagonal |
| Original Image | Y | 0.4499642610 | 0.3973681636 | 0.8772539012 |
| | Cb | 0.6136495805 | 0.5500217651 | 0.6312938787 |
| | Cr | 0.6737330303 | 0.6218992932 | 0.6852202553 |
| Encrypted Image | Y | 0.0387610216 | 0.0989613447 | 0.0973557834 |
| | Cb | 0.0465625961 | 0.0942653695 | 0.0157075666 |
| | Cr | 0.0534913613 | 0.0565025022 | -0.0207536557 |
| GIF Images | | Orientation | | |
| | | Horizontal | Vertical | Diagonal |
| Original Image | Red | 0.9103363915 | 0.9236943997 | 0.9370063144 |
| | Green | 0.9271147209 | 0.9343745458 | 0.9432646400 |
| | Blue | 0.8133623154 | 0.8304941063 | 0.8354562874 |
| Encrypted Image | Red | 0.0021957958 | 0.0027551506 | 0.0114763303 |
| | Green | 0.0042820322 | 0.0035947146 | 0.0135946511 |
| | Blue | 0.0024500781 | 0.0048780316 | 0.0110798046 |

The results confirm that our solution reduces the correlation between adjacent pixels of the image distinctly.

## 4.3 Sensitivity analysis

(1) Sensitivity analysis of the initial values

The test result is displayed as Fig. 8 and Fig. 9. The correct initial values are -0.1790288311 and -0.1628589871. The wrong initial values for decryption are -0.17902883110001 and -0.1628589871, with a small fluctuation to the first parameter.

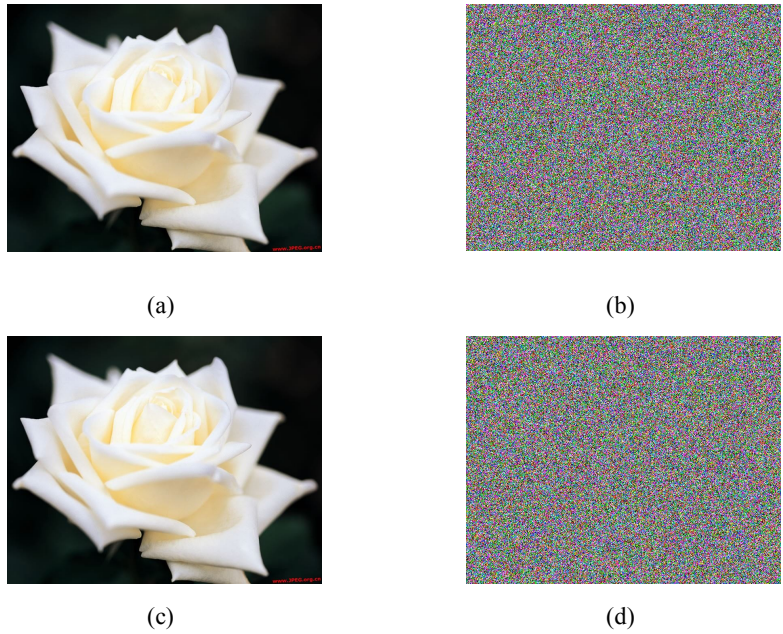

Fig. 8 JPEG images (a) original image, (b) encrypted image, (c) decrypted image and (d) decrypted image with wrong initial values

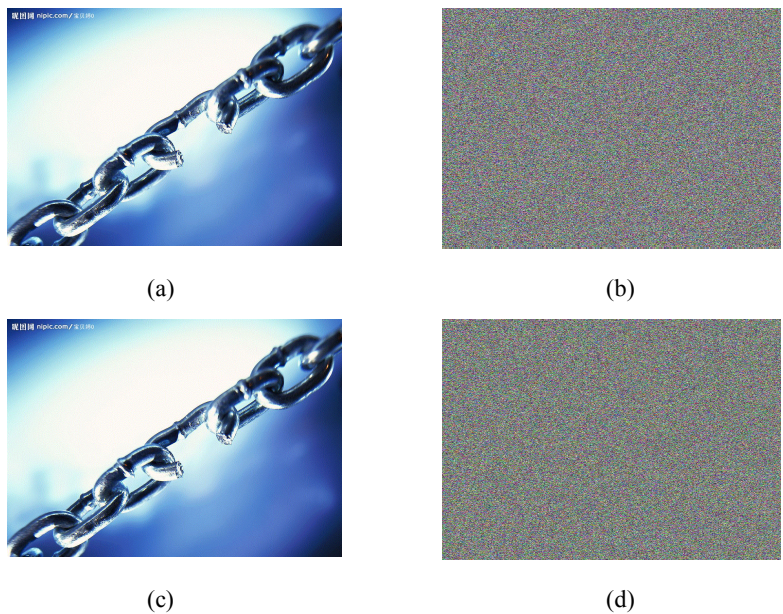

Fig. 9 GIF images (a) original image, (b) encrypted image, (c) decrypted image and (d) decrypted image with wrong initial values

Obviously, a minor distance ($10^{-14}$) between the initial values will lead to the failure of the decryption.

(2) Sensitivity analysis of the plaintext

The test result is displayed as Table 2. The initial values are -0.1790288311 and -0.1628589871.

The keys for 3DES and AES128 are both "123321", a string with length of 6.

Table 2 Sensitivity to plaintext

| | JPEG | | New scheme | 3DES | AES128 |
|---|---|---|---|---|---|
| Sensitivity test (1 bit difference in plaintext) | NPCR | Y | 0.9921895441 | 2.04e-5 | 3.88e-5 |
| | | Cb | 0.9922324699 | 7.36e-5 | 1.47e-4 |
| | | Cr | 0.9918972397 | 7.36e-5 | 1.47e-4 |
| | UACI | Y | 0.3361135490 | 7.76e-5 | 1.40e-5 |
| | | Cb | 0.3360163342 | 2.82e-5 | 4.68e-5 |
| | | Cr | 0.3337855433 | 2.82e-5 | 4.68e-5 |
| | GIF | | New scheme | 3DES | AES128 |
| Sensitivity test (1 bit difference in plaintext) | NPCR | Red | 0.9911632357 | 1.06e-5 | 1.99e-5 |
| | | Green | 0.9927903982 | 1.06e-5 | 1.99e-5 |
| | | Blue | 0.9921649741 | 1.06e-5 | 1.99e-5 |
| | UACI | Red | 0.3341026215 | 2.84e-6 | 6.73e-6 |
| | | Green | 0.3345633391 | 3.25e-6 | 7.19e-6 |
| | | Blue | 0.3421481330 | 3.33e-6 | 4.93e-6 |

From the test results, we can conclude that our new scheme has a satisfied performance on sensitivity to plaintext. It is much better than that of the classical symmetrical cryptographic methods.

(3) Avalanche effect

The test result is displayed as Table 3. The initial values are -0.1790288311 and -0.1628589871.

Table 3 Avalanche Effect

| Avalanche criterion | New scheme (%) | AES128 (%) | 3DES (%) |
|---|---|---|---|
| JPEG | 49.96 | 7.01e-05 | 3.39e-05 |
| GIF | 49.81 | 1.04e-05 | 5.41e-06 |

As you can see, our proposed scheme even obtains a better performance on avalanche effect than the traditional solutions.

4.4 Information entropy analysis

Approximate entropy of a random sequence is as follows.

$$H(S) = \sum_{S} P(S_i) \log_2 \frac{1}{P(S_i)} \text{ bits} \tag{16}$$

In Formula (16), $P(S_i)$ denotes the probability of symbol $S_i$. For calculating the approximate entropy, the probability can be replaced by the ratio of occurrences of the specified symbol to the total length of the random sequence.

The test result is displayed as Table 4. The initial values are -0.1790288311 and -0.1628589871.

Table 4 Entropy test

| JPEG | Entropy | |
|---|---|---|
| Cipher entropy (theoretical value is 7) | Y | 6.9998167604 |
| | Cb | 6.9990590012 |
| | Cr | 6.9993319890 |
| GIF | Entropy | |
| Cipher entropy (theoretical value for total is 8) | Red | 7.0453960363 |
| | Green | 7.2529295091 |
| | Blue | 7.1479725253 |
| | Total | 7.9997870729 |

As most of the values of plaintext ranges from -128 to 127, the ideal value of the ciphered entropy of JPEG file should be 7, since we have to only modify the least significant 7 bits as it is not necessary to encrypt the higher bits. Therefore, the test results above are acceptable.

Note that the total number of colors that are allowed to be used in an individual GIF image is limited. Therefore, the approximate entropies of red, green and blue components are not near 8, which is the ideal entropy value for full color (32 bits) ciphered bitmaps.

4.5 NIST SP 800-22 test results of cipher

The test results are shown in Table 5 and Table 6. The initial values are -0.1790288311 and -0.1628589871 for JPEG, and they are 0.1 and 0.2 for GIF.

Table 5 NIST SP 800-22 test results of JPEG ciphered image

| Statistical test | P-value | Result |
|---|---|---|
| Frequency | 0.6925971712 | Success |
| Block Frequency (m=128) | 0.2730042852 | Success |
| Runs | 0.5326579147 | Success |
| Long runs of ones (M=10000, N=75) | 0.4467952762 | Success |
| Rank | 0.9045338866 | Success |
| Spectral DFT | 0.2853057611 | Success |
| Non-overlapping templates (m=9, B=101001100) | 0.5952406178 | Success |
| Overlapping templates (m=9, M=1032, N=968) | 0.0182820069 | Success |
| Universal (L=7, Q=1280, K=141577) | 0.8190968429 | Success |
| Lempel Ziv complexity | 0.7509607391 | Success |
| Linear complexity (M=1000) | 0.2336958734 | Success |
| Serial | | |
| p-value1 | 0.9233773251 | Success |
| p-value2 | 0.8336811505 | Success |
| Approximate entropy (m=10) | 0.1292246561 | Success |
| Cumulative sums | | |
| Forward | 0.7504396563 | Success |
| Reverse | 0. 1935254961 | Success |
| Random excursions (state x) | | |
| x=-4 | 0.0593995871 | Success |
| x=-3 | 0.7032899184 | Success |
| x=-2 | 0.4077410179 | Success |
| x=-1 | 0.0231394303 | Success |
| x=1 | 0.3480722776 | Success |
| x=2 | 0.3376947711 | Success |
| x=3 | 0.8643339969 | Success |
| x=4 | 0.7291568619 | Success |
| Random excursions variant (state x) | | |
| x=-9 | 0.4646998509 | Success |
| x=-8 | 0.0971460791 | Success |
| x=-7 | 0.1565194142 | Success |
| x=-6 | 0. 2040135903 | Success |
| x=-5 | 0. 2345890619 | Success |
| x=-4 | 0.1808529572 | Success |
| x=-3 | 0.1456615720 | Success |
| x=-2 | 0.1534593283 | Success |
| x=-1 | 0.1615329445 | Success |
| x=1 | 0.6078998587 | Success |
| x=2 | 0.7111746529 | Success |
| x=3 | 0.5168516766 | Success |
| x=4 | 0.6838320773 | Success |
| x=5 | 0.8911733760 | Success |
| x=6 | 0.9445009924 | Success |
| x=7 | 0.7569376482 | Success |
| x=8 | 0.8658708484 | Success |
| x=9 | 0.8739373537 | Success |

Table 6 NIST SP 800-22 test results of GIF ciphered image

| Statistical test | P-value | Result |
|---|---|---|
| Frequency | 0.4269499765 | Success |
| Block Frequency (m=128) | 0.9374763966 | Success |
| Runs | 0.0830653238 | Success |
| Long runs of ones (M=10000, N=75) | 0.8612727713 | Success |
| Rank | 0.7295034038 | Success |
| Spectral DFT | 0.0770114610 | Success |
| Non-overlapping templates | 0.8800811759 | Success |
| (m=9, B=101001100) | | |
| Overlapping templates (m=9, M=1032, N=968) | 0.0124741817 | Success |
| Universal (L=7, Q=1280, K=141577) | 0.0750912920 | Success |
| Lempel Ziv complexity | 0.0124236313 | Success |
| Linear complexity (M=1000) | 0.2413868379 | Success |
| Serial | | |
| p-value1 | 0.3613836187 | Success |
| p-value2 | 0.7000373804 | Success |
| Approximate entropy (m=10) | 0.2479288348 | Success |
| Cumulative sums | | |
| Forward | 0.7693609544 | Success |
| Reverse | 0.1560653149 | Success |
| Random excursions (state x) | | |
| x=-4 | 0.1235959934 | Success |
| x=-3 | 0.1365390701 | Success |
| x=-2 | 0.2479408135 | Success |
| x=-1 | 0.1250459204 | Success |
| x=1 | 0.0635319405 | Success |
| x=2 | 0.9240437593 | Success |
| x=3 | 0.9915710212 | Success |
| x=4 | 0.8075050539 | Success |
| Random excursions variant (state x) | | |
| x=-9 | 0.0269201835 | Success |
| x=-8 | 0.1655410584 | Success |
| x=-7 | 0.1759873588 | Success |
| x=-6 | 0.2215934597 | Success |
| x=-5 | 0.4233650689 | Success |
| x=-4 | 0.4654701339 | Success |
| x=-3 | 0.4691678089 | Success |
| x=-2 | 0.3794348376 | Success |
| x=-1 | 0.2893521693 | Success |
| x=1 | 0.3347960200 | Success |
| x=2 | 0.9825776662 | Success |
| x=3 | 0.7736851609 | Success |
| x=4 | 0.6576400218 | Success |
| x=5 | 0.5964387877 | Success |
| x=6 | 0.6198358674 | Success |
| x=7 | 0.7649592671 | Success |
| x=8 | 0.7361724793 | Success |
| x=9 | 0.6864805904 | Success |

All required entries have been passed. The ciphered image has a good randomness.

4.6 Encryption speed test

The test results are displayed in Table 7. It is inspiring that our proposed scheme is totally faster than traditional symmetrical cryptographic solutions. This is because the iteration of chaotic cryptography is easier to compute than traditional approaches, such as 3DES and AES128.

Table 7 Encryption speed test results

| Image size | Time (s) | | |
|---|---|---|---|
| | New scheme | 3DES | AES128 |
| House. jpg 256*256 | 1.263 | 2.652 | 7.911 |
| House. jpg 512*512 | 5.086 | 10.595 | 31.933 |
| House. jpg 1024*1024 | 20.384 | 41.772 | 127.437 |
| House. gif 256*256 | 0.873 | 1.841 | 6.131 |
| House. gif 512*512 | 3.354 | 7.425 | 23.946 |
| House. gif 1024*1024 | 13.245 | 29.203 | 96.112 |

Our scheme has an obvious advantage on computational speed towards other classical solutions, like 3DES and AES128.

## 5. Conclusion

New JPEG and GIF oriented compound chaotic image encryption schemes, which are combined with 3D baker as permutation, have been proposed. Its computational speed is faster than the traditional solutions of chaotic image encryption. The results of security tests affirm that our proposed approach is sensitive to both the key and the plaintext, and the ciphered images own good randomness. The security of our new scheme is thus verified.

## 6. Acknowledgement

This work was supported by the National Natural Science Foundation of China (Grant No. 60973162), the Natural Science Foundation of Shandong Province (Grant No. ZR2009GM037), the Science and technology of Shandong Province of China (Grant No. 2010GGX10132), the Science and technology of Shandong Province of China (Grant No. No.2012GGX10110 ）, the Scientific Research Foundation of Harbin Institute of Technology at Weihai (Grant No. HIT (WH) ZB200909), and the Key Natural Science Foundation of Shandong Province of China (Grant No. Z2006G01).